\documentclass[10pt,letterpaper,twocolumn]{article} 

\usepackage{ol2}
\usepackage[draft]{hyperref}
\usepackage{amsmath}
\newcommand {\startbild}{\begin{figure}}
\newcommand {\stopbild}{\end{figure}}
\newcommand {\staf}{\begin{equation}}
\newcommand {\stof}{\end{equation}}
\newcommand {\staffeld}{\begin{eqnarray}}
\newcommand {\stoffeld}{\end{eqnarray}}
\newcommand {\staa}{\begin{align}}
\newcommand {\stoa}{\end{align}}

\newcommand{\ket}[1]{|#1\rangle}
\newcommand{\bra}[1]{\langle #1|}
\renewcommand{\vec}[1]{{\bf #1}}
\begin{document}
\twocolumn[
\title{Quantum interference and entanglement of photons which do not overlap in time}

\author{R. Wiegner,$^{1,*}$ C. Thiel,$^{1}$ J. von Zanthier,$^{1}$ and G. S. Agarwal$^{2}$}

\address{
$^1$Institut f\"ur Optik, Information und Photonik, Universit\"at Erlangen-N\"urnberg, Erlangen, Germany
\\
$^2$Department of Physics, Oklahoma State University, Stillwater, OK, USA \\
$^*$Corresponding author: ralph.wiegner@physik.uni-erlangen.de
}

\date{\today}

\begin{abstract}We discuss the possibility of quantum interferences and entanglement of photons which exist at different intervals of time, i.e., one photon being recorded before the other has been created. The corresponding two-photon correlation function is shown to violate Bell's inequalities.\end{abstract}

\maketitle
]

Coherent superposition of states, quantum interference and entanglement are at the heart of quantum physics, playing a key role in fundamental investigations as well as potential applications.  So far, the prototypical experimental setup to observe higher order quantum interferences consists in photon pairs, produced by parametric down conversion (PDC) in nonlinear crystals, subsequently sent onto a beam splitter \cite{HOM87,Shih88}. In this way Hong-Ou-Mandel (HOM) two-photon interferences \cite{HOM87} have been observed with the two photons being identical in their spectral, spatial and polarization modes and measured within their coherence time \cite{Zukowski93}. HOM two-photon interferences have been demonstrated also with uncorrelated photons. These may stem from independent or even disparate sources like two PDC sources \cite{Zeilinger06}, two atoms, ions, semiconductor quantum dots \cite{Grangier06}, or a quantum dot or PDC source interfering with a coherent or thermal source \cite{Franson03}.

The temporal overlap of the individual photons at a beam splitter is not a fundamental requirement for the observation of two-photon quantum interferences, the ultimate constraint being rather the indistinguishability of the corresponding two-photon amplitudes \cite{Chiao92,Shih96,Franson89}.
For example, Franson proposed a two-photon interferometer where two photons, simultaneously emitted from a two-photon source, propagate towards the detectors along spatially separated paths \cite{Franson89}. Since it is assumed that the source emits the two photons at the same time but leaves the exact moment of emission unpredictable, coincident detection as well as identical optical path lengths for the individual photon paths are required in order to obtain indistinguishability of the two emerging two-photon amplitudes. These give rise to quantum interferences even though the time delay between them may be much larger than the individual single photon (first order) coherence time \cite{Franson89}. 

In this paper we discuss a different two-photon interferometer where the photons are emitted from independent single photon sources (SPS) \cite{Mandel83}. By recording the photons in the far field and by assuming that each detector records precisely one photon, two two-photon amplitudes appear which differ in phase and contribute both coherently to the two-photon signal \cite{Skornia01}. As we assume uncorrelated emitters, here no assumption with respect to the time of emission of the individual photons can be formulated. The question then arises what requirements with respect to the emission time and time of detection of the two photons have to be fulfilled in order to obtain indistinguishability between the two-photon amplitudes and to observe two-photon interferences. 

In the following we demonstrate that for identical SPSs in the considered interferometer no such requirements exist. In particular, we show that the individual optical paths of the two photons may not only differ by an amount larger than their individual coherence lengths but that the time delay between the detection of the two photons may even be larger than the transit time of the photons from the source towards the detectors. This is equivalent to the statement that two-photon interferences may be observed even though the two photons exist at different intervals of time, i. e., one photon being recorded before the other has been emitted. We also demonstrate that this two-photon signal may violate Bell's inequalities 
 in the Clauser and Horne (CH74) formulation \cite{CH74}. In view of the above statement this implies that entanglement among two photons may exist even though the two photons do not overlap in time.

\begin{figure}[h!]
\centering
\includegraphics[width=0.4\textwidth]{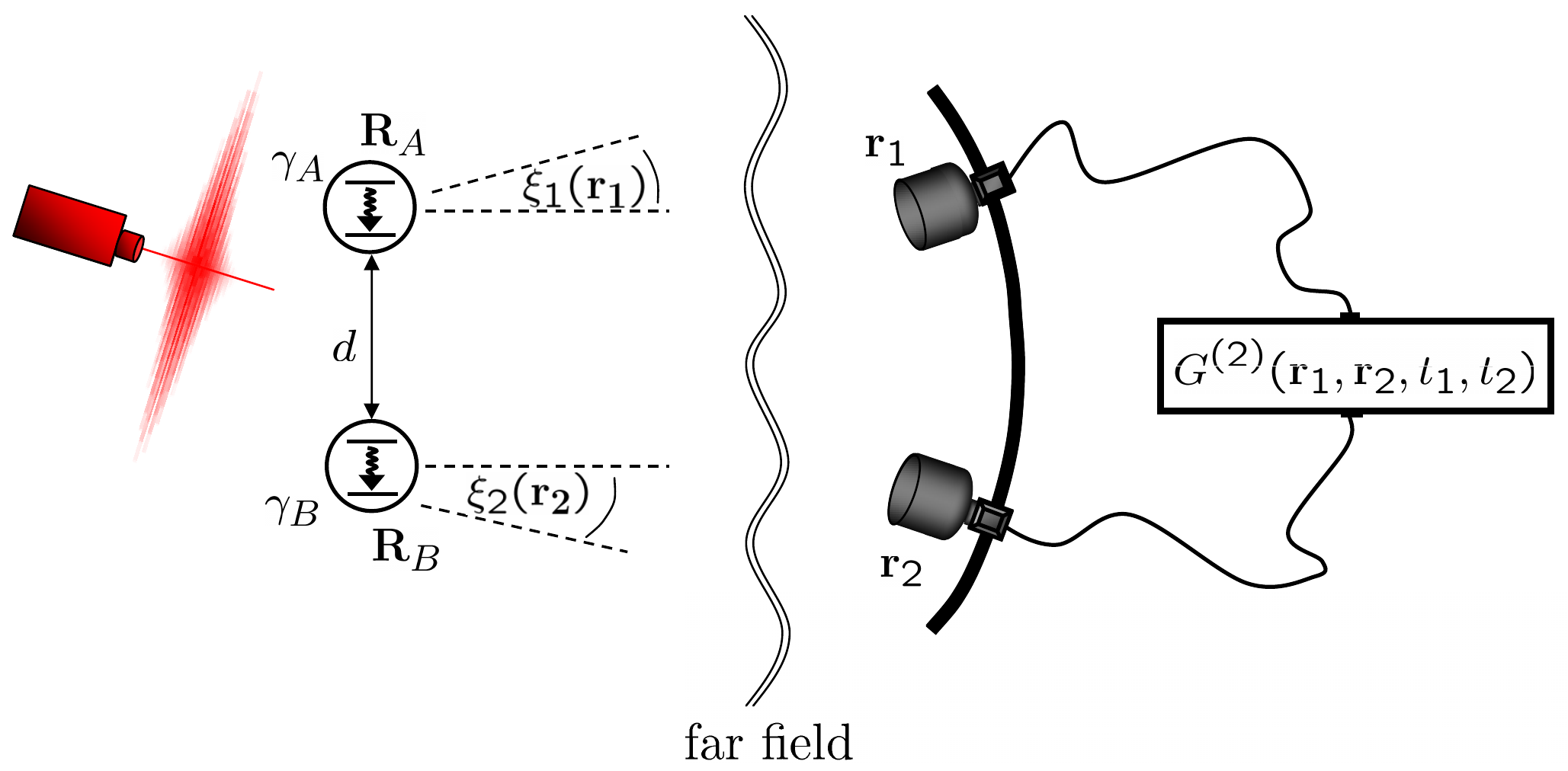}
\caption{\label{setup} 
Two photons are emitted by two independent single photon sources (SPS) A and B, located at ${\bf R}_A$ and ${\bf R}_B$, with decay constants $\gamma_A$ and $\gamma_B$, respectively. The two photons are measured at times $t_i$ by two detectors at ${\bf r}_i$ (i = 1,2) in the far field of the SPSs.}
\end{figure}

Let us consider a system of two independent SPSs which upon excitation each emit a single photon at a random time by the process of spontaneous decay; the exact functional dependence of the decay process is however of no importance. Two detectors measure the photons in the far field of the SPSs so that they can not distinguish from which of the two SPSs the photons were emitted. We define a measurement cycle by two detection events at the two detectors, i.e., we assume that each of the two detectors registers precisely one photon so that two photon absorption processes at one detector are excluded. 

As an example for two independent SPSs, we consider two uncorrelated and potentially disparate two-level atoms $A$ and $B$, with lower states $|g\rangle_A$ ($|g\rangle_B$), upper states $|e\rangle_A$ ($|e\rangle_B$) and decay constants $\gamma_A$ ($\gamma_B$), where the corresponding transition frequencies $\omega_A = \omega_B = \omega$ are assumed to be identical. The two atoms are localized at positions ${\bf R}_A$ and ${\bf R}_B$ with a separation $d >\!> \lambda$, so that any interaction between them can be neglected. The two detectors are located at positions ${\bf r}_1$ and ${\bf r}_2$, where $|\vec{r}_i - \vec{R}_n| \gg d$ ($i = 1,2$, $n = A, B$) so that the far field condition is fulfilled (see ~Fig.~\ref{setup}).

Let $\hat{{\bf E}}^{(+)} ({\bf r}_i, t_i)$ be the positive frequency part of the total electric field operator at position ${\bf r}_i$ and time $t_i$ (i = 1,2), given by
\begin{eqnarray}
\label{tutg2}
\hat{{\bf E}}^{(+)} ({\bf r}_i, t_i) \equiv \hat{{\bf A}}^{(+)}_i +  \hat{{\bf B}}^{(+)}_{i} \hspace{2cm}\nonumber \\
 = E_A \, e^{-ik({\bf \hat{r}}_i\cdot{\bf R}_A)}\,\hat{S}^+_A(\tilde{t}_i)+ E_B \, e^{-ik({\bf \hat{r}}_i\,{\bf R}_B)}\,\hat{S}^+_B(\tilde{t}_i)\;.
\end{eqnarray}
\noindent Here, $k=\frac{2\pi}{\lambda}=\frac{\omega}{c}$ is the wave number of the scattered photons, ${\bf \hat{r}}_i:=\frac{{\bf r}_i}{|{\bf r}_i|}$ is a unit vector in the direction of the $i$th detector, and the amplitudes of the electric fields are given by $E_A = E_B = \sqrt{\hbar\,\omega/\epsilon_0\,V}$. The operator $\hat{S}^+_n(\tilde{t}_i)$ corresponds to the time dependent lowering operator $\ket{g}_n\bra{e}$ of the transition $|e\rangle_n \rightarrow |g\rangle_n$ in the Heisenberg picture for the two-level atom $n$ ($n = A, B$). We denote with $t_i$  the (retarded) detection time of a photon at ${\bf r}_i$ and with $\tilde{t}_i = t_i - t_{R_i}$ the emission time, where $t_{R_i} = \frac{r_i}{c}$ is the time of propagation from the atoms towards the detector at ${\bf r}_i$.

Using the master equation for the density matrix $\rho$ of a two-level atom in case of pulsed excitation 
\staffeld
\label{MEQ}
\frac{\partial \rho_n}{\partial t} = - i \omega \left[ \hat{S}_n^z, \rho \right] - \gamma \left( \hat{S}_n^+ \hat{S}_n^- \rho - 2  \hat{S}_n^- \rho \hat{S}_n^+  +  \rho \hat{S}_n^- \hat{S}_n^+  \right),\nonumber
\stoffeld
and employing the quantum regression theorem 
, we obtain for the two-time expectation values $\langle \hat{S}_n^{-} (t_j) \hat{S}_n^{+} (t_i)\rangle$ 
\staffeld
\label{timemoments2}
\langle \hat{S}_n^{-} (t_j) \hat{S}_n^{+} (t_i)\rangle &=& e^{(i \omega - \gamma_n)(t_j-t_i)} \langle \hat{S}_n^{-} (t_i) \hat{S}_n^{+} (t_i)\rangle \nonumber\\
&=& e^{i \omega (t_j-t_i) - \gamma_n (t_j+t_i)} \langle \hat{S}_n^{-} (0) \hat{S}_n^{+} (0)\rangle\;,\nonumber
\stoffeld
with $t_j > t_i$ ($i, j = 1,2$, $n = A, B$).
In Glauber notation 
 and with the foregoing considerations
the second order correlation function can be written as 

\begin{eqnarray}
\label{g2times}
&&G^{(2)}({\bf r}_1,{\bf r}_2,t_1,t_2) \hspace{2cm}\nonumber\\
&&=  \langle \hat{{\bf E}}^{(-)}\!({\bf r}_1, t_1)\hat{{\bf E}}^{(-)}\!({\bf r}_2, t_2)\hat{{\bf E}}^{(+)}\!({\bf r}_2, t_2)\hat{{\bf E}}^{(+)}\!({\bf r}_1, t_1)\rangle \nonumber\\
&&= 2\, E_A^4 \left( e^{- 2 (\gamma_A\,\tilde{t}_1 + \gamma_B\,\tilde{t}_2 )} \right.\nonumber\\
&&\hspace{2cm}\left. + e^{- (\gamma_A + \gamma_B)(\tilde{t}_1 + \tilde{t}_2)}  \cos{\left[\varphi_2-\varphi_1\right]}\right) \;,
\end{eqnarray}
with $(\hat{{\bf E}}^{(-)} )^\dagger = \hat{{\bf E}}^{(+)}$ and the relative phase $\varphi_i$ given by $\varphi_i \equiv \varphi({\bf r}_i)= d \, k\,\sin[\xi(\vec{r}_i)]$, where $\xi(\vec{r}_i)$ is indicated in~Fig.~\ref{setup}. 

\noindent In the far field, with $|\vec{r}_i - \vec{R}_n| \gg d$ and  assuming for simplicity $t_{R_1} \approx t_{R_2} \approx t_{R}$, we obtain from Eq.~(\ref{g2times}) for the visibility \mbox{${\cal V}$} of the two-photon correlation signal 

\begin{eqnarray}
\label{vis2}
{\cal V}(t_1,t_2) & := &\frac{G^{(2)}_{max} - G^{(2)}_{min}}{G^{(2)}_{max} + G^{(2)}_{min}}
=\frac{e^{- (\gamma_A + \gamma_B)(t_1 + t_2 )}}{e^{- 2 \gamma_A t_1 - 2 \gamma_B t_2}} \nonumber \\
&& \xrightarrow{\gamma_A = \gamma_B = \gamma} \; {\cal V}_\gamma = 1 \,.
\end{eqnarray}

\begin{figure}[h!]
\centering
\includegraphics[width=0.4 \textwidth]{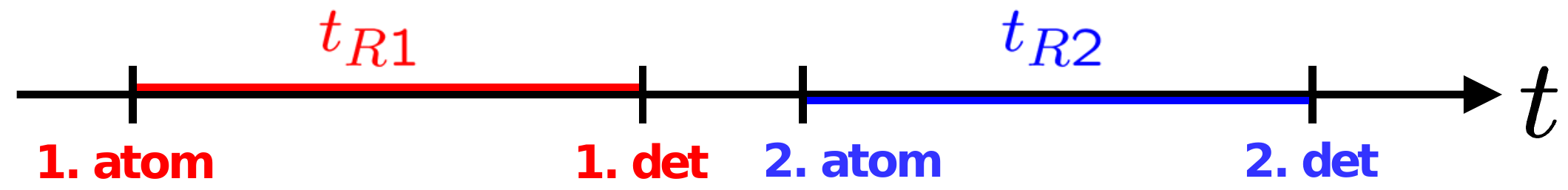}
\caption{\label{scen21} (color online) Schematic representation of the scenario that the first scattered photon is detected before the second photon is emitted. In this case the two photons do not exist in the same interval of time.}
\end{figure}

\noindent From Eq.~(\ref{vis2}) we can see that if the two decay constants $\gamma_A$ and $\gamma_B$ differ significantly, coincident detection of the two photons, $t_1 \approx t_2$, is required to obtain a substantial visibility. However, if the two decay constants are equal, $\gamma_A = \gamma_B = \gamma$, the visibility of the two-photon correlation signal (abbreviated ${\cal V}_\gamma$) becomes independent of the detection times of the two photons as we simply obtain ${\cal V}_\gamma = 1$. 
This means that the time difference $t_2 - t_1$ between the two detection events may not only exceed the coherence time $\tau_c \approx 1/\gamma$ of the individual photons but in particular it may exceed the time a photon needs to propagate from the atoms towards the detector. This allows for the following scenario: upon excitation, an atom scatters a single photon after about the decay time $\gamma^{-1}$. Without loss of generality the photon propagates from the atoms towards the first detector in a time $t_{R}$ and is measured in the far field at position ${\bf r}_1$ and time $t_{1}$. Due to the probabilistic nature of spontaneous decay the other atom may scatter \emph{thereafter} a photon that is measured after the propagation time $t_{R}$ by the second detector at time $t_{2}$ at ${\bf r}_2$ (see Fig.~\ref{scen21}). Taking into account only measurements where the time delay between the two detection events $t_{2} - t_{1}$ exceeds $t_{R}$, we thus obtain a  two-photon interference signal with a visibility ${\cal V}_\gamma = 100\%$ among two photons which never existed in the same interval of time.

We can appreciate the situation by investigating the specific atomic state produced after the measurement of the first photon and before the measurement of the second photon. It can be written in the form
 \staf
\Psi({\bf r}_1) = \frac{1}{\sqrt{2}} \left( \ket{g\;e} + e^{i\phi({\bf r}_1)} \ket{e\;g} \right) \;,
\stof

\noindent which expresses the absence of information about the particular atom emitting the photon due to the far field detection scheme: either the first atom has scattered the photon and is transfered to the ground state or the second atom has emitted the photon; both possibilities have to be taken into account to fully describe the state of the system, whereby the two possibilities differ by a phase factor $e^{i\phi({\bf r}_1)}$. The state $\Psi({\bf r}_1)$ thus acts as a \emph{quantum memory} which stores the information about the phase difference $\phi({\bf r}_1)$ between these two possibilities - even though the second photon has not yet been emitted. Starting from the state $\Psi({\bf r}_1)$ the  second photon is then scattered. 

By post-selecting only those detection events where the second photon is measured by the second detector at a time $t_2 > t_R + t_1$, a second phase difference $\phi({\bf r}_2)$ appears between the two possibilities that either the second or the first atom has scattered the second photon; both phase differences, $\phi({\bf r}_1)$ as well as $\phi({\bf r}_2)$, then contribute to the total photon-photon correlation signal. Hereby, the two-photon amplitudes remain indistinguishable at any time due to the far-field detection scheme and the probabilistic nature of the decay process: each two-photon amplitude consists of two - delayed - photons of which the precise origin is unknown, i.e., no \emph{which-way} information can be obtained, even if one photon is already measured and the second photon is not yet detected.


Note that for $\gamma_A = \gamma_B = \gamma$ the normalized second order correlation function takes the form (cf.~Eq.~(\ref{g2times})) 
\begin{eqnarray}
\label{g2timessimple}
g^{(2)}(\vec{r}_1, \vec{r}_2,t_1,t_2) = \frac{1}{2}\,e^{- 2 \gamma(\tilde{t}_1 + \tilde{t}_2)} ( 1 + \cos{\left[\varphi_2-\varphi_1\right]})\,.
\end{eqnarray}

\noindent Inserting this expression into the Bell inequalities in CH74 formulation \cite{CH74} the inequalities read \cite{wiegner1}
\staffeld
\label{ch1}
- e^{- 2 \gamma (\tilde{t}_{10} + \tilde{t}_{20}) }\leq g^{(2)}(1,2)- g^{(2)}(1,2')  + g^{(2)}(1',2) \nonumber\\
+ g^{(2)}(1',2')- e^{- 2 \gamma (\tilde{t}_1' +\tilde{t}_{20}) }- e^{- 2 \gamma (\tilde{t}_{10} +\tilde{t}_{2}) }\leq 0 ,\hspace{1cm}
\stoffeld
\noindent for ${t}_{i0} \leq {t}_{j}$ and with $g^{(2)}(i,j) \equiv g^{(2)}(\vec{r}_i, \vec{r}_j,t_i,t_j)$. By post-selecting detection events where we have a \emph{fixed} time delay $t_{2} - t_{1}$ and $t_i = t_{i'} = t_{i0}$ the time dependency cancels. 
If in addition we choose \mbox{$t_{2} - t_{1} > t_{R}$}, we arrive with the Bell angles ($\frac{\pi}{4}, \frac{3\pi}{4}, \frac{\pi}{4}, \frac{\pi}{4}$) for the relative detector positions $\varphi_i - \varphi_j$ at a violation of the CH74 inequalities, i.e., at an entanglement between photons which did not exist in the same interval of time\footnote{Note that to exclude communication between the two detectors during the measurement of the two photons the condition $2\,t_R > t_2 - t_1 > t_R$ must be satisfied. This could be fulfilled by placing the two detectors opposite to each other with respect to the two photon sources.  For example, for 
trapped $Ca^+$ ions the natural linewidth of the D1 transition is $\gamma = 20 MHz$ and the lifetime $(2 \pi \gamma)^-1 = 8 ns$. If the distance between the ions and the detectors is $r = 1\,m$, then $t_2 - t_1 > t_R = r/c = 3 ns$ can be easily fulfilled.}.

In conclusion we studied quantum interferences and violation of Bell's inequalities with photons from independent single photon sources which do not overlap in time. We showed that such interferences can produce a visibility of 100\% and violate Bell's inequalities in the CH74 formulation. The appearance of quantum interferences can be explained by the lack of spatial and temporal information about the photons in the considered interferometer: due to the far field detection scheme and the probabilistic nature of the emission process indistinguishability of the two-photon amplitudes is preserved, independent of the photon detection times. As specific sources we considered independent atoms. However, other single photon sources will display a similiar behavior, e.g., trapped ions, quantum dots or single molecules.

RW gratefully acknowledges financial support by the Elite Network of Bavaria and the hospitality at the Oklahoma State University in spring 2010 and 2011. This work was supported by the DFG.

\end{document}